\documentclass[twocolumn,trackchanges]{aastex631}

\shorttitle{Gamma-ray QPO in blazar S2~0109+22}
\shortauthors{Zhang et al.}
\graphicspath{{figures/}}
\usepackage{amsmath}
\begin{document}
\title{The detection of possible $\gamma$-ray quasi-periodic modulation with $\sim$600 days from the blazar S2~0109+22}
\author[0000-0003-3392-320X]{Haoyang Zhang}
\affiliation{Department of Astronomy, Key Laboratory of Astroparticle Physics of Yunnan Province, Yunnan University, Kunming 650091, China}
\author[0000-0002-6292-057X]{Fan Wu}
\affiliation{Department of Astronomy, Key Laboratory of Astroparticle Physics of Yunnan Province, Yunnan University, Kunming 650091, China}
\author[0000-0001-7908-4996]{Benzhong Dai}
\correspondingauthor{Benzhong Dai}{\email{bzhdai@ynu.edu.cn}}
\affiliation{Department of Astronomy, Key Laboratory of Astroparticle Physics of Yunnan Province, Yunnan University, Kunming 650091, China}

\begin{abstract}
In this work, we analyzed the long-term $\gamma$-ray data by a Fermi Large Area Telescope (Fermi-LAT) of blazar S2~0109+22, ranging from 2008 to 2023. The quasi-periodic oscillations (QPOs) of blazars aided in investigating the physical properties of internal supermassive black holes, the nature of variability, and the underlying radiation mechanism. We employed four different methods---Weighted Wavelet Z-transform, Lomb-Scargle periodogram, REDFIT and phase folded light curve analysis, for searching QPO signals. Our analysis identified a possible QPO behavior with a periodicity of $\sim600$ days in November 2013 to January 2023 at a significance level of $\sim3.5\sigma$. This QPO signal sustained $\sim9$ years, corresponding to 5.6 cycles, which was in good agreement with the previously observed of periodicity $\sim657$ days in radio. We explained this phenomenon based on the accretion model and the lighthouse effect, in a binary black hole system.
\end{abstract}

\section{Introduction} \label{sec:intro}
Active galactic nucleis (AGN) emits extreme non-thermal and thermal radiation across all-wavelengths, from radio to high energy $\gamma$-rays. AGNs are generally considered as accretion systems with a supermassive black hole (SMBH) at their centre, providing the necessary power \citep{2017A&ARv..25....2P}. AGNs can be classified into jetted and non-jetted \citep{2017NatAs...1E.194P}. When the jet is oriented close to our line of sight, it form the subclass called the blazars \citep{1995PASP..107..803U}. Blazars are classified into BL Lacertae objects (BL Lacs) and flat spectrum radio quasars (FSRQs). Generally, blazars always have a variability in all-wavelengths with different timescales \citep{1995ARA&A..33..163W,2001AJ....122.2901D,2009MNRAS.392.1181D,2015ApJS..218...18D}.  According to the variability timescale, the variability is categorized into intraday variability with the timescales of a single day, the short-term variability with timescales of a few days to months, and the long-term variability with timescales of years \citep{2011IAUS..275..164F,2021ApJS..253...10F,2022ApJS..260...47C}.

With the increase of observational data, a special variability phenomenon of quasi-periodic oscillations (QPOs) has been discovered in AGNs across all energy spectrum. For non-jetted AGNs, which comprise the majority of existing AGNs, quasi-periodic behavior has been observed in both optical and X-ray emissions. This quasi-periodic behavior has been associated with the accretion disk emission and usually occurs over the time scales of hours to months, such as $\sim$ 44 days in KIC 9650712 \citep{2018ApJ...860L..10S}, $\sim$1 hour in RE J1034+396 \citep{2008Natur.455..369G}, $\sim$ 2 hours in Mrk 766 \citep{2017ApJ...849....9Z}. Harmonic oscillations with the ratios of 1:2 and 1:3 have been detected in 1H 0707-495 and Mrk142, respectively \citep{2018ApJ...853..193Z,2021MNRAS.502.1158Z}. These QPOs can often be interpreted as models of orbital resonances \citep{2003PASJ...55..467A,2008A&A...486....1H,2015ApJ...798L...5Z}. In this scenario, the states of particles along different directions on the accretion disk (i.e., orbital, vertical, and radial epicyclic) are coupled to create resonance. A doubt about the applicability of this model to AGNs has been raised by \citet{2021ApJ...906...92S}. There are certain alternative models that have been proposed, such as the seismic models \citep{1997ApJ...476..589P,2003MNRAS.344L..37R}, the accretion disk hot spot models \citep{2004AIPC..714...40S}, and a transient chaos accretion model called the “Dripping Handrail” \citep{1993ApJ...411L..91S,1996ApJ...468..617Y}. Therefore, studying the QPO behavior in non-jetted AGNs is crucial for understanding the physical structure of SMBH accretion systems. Further observational data is needed to support the possible physical explanations.

The jetted AGNs exhibit strong non-thermal radiation from relativistic jets. In the radio band of 15~GHz, the periodic signals from a few hundred days to several years have been reported, i.e., $\sim$150 days in J1359+4011, $\sim$270 days for blazar PKS~0219-164, 4.69 years in PKS~J2134-0153, and 965 days  for AO~0235+164 (14.5 GHz) \citep{2013MNRAS.436L.114K, 2017ApJ...847....7B,2021MNRAS.506.3791R,2021MNRAS.501.5997T}.
In the optical, OJ 287 has been identified as the most likely candidate for a binary black hole system with a 12 years period \citep{1985A&A...147...67S,1988ApJ...325..628S,2006ApJ...646...36V,2008Natur.452..851V}, and certain shorter periodic signals of $\sim$400 and $\sim$800 days have been reported in OJ 287 \citep{2016ApJ...832...47B}. In addition, the periodic signals of $\sim$317 days, $\sim$283 days and 6.4 years for PKS 2155-304, 1823+568 and SDSS~J0752, have been reported \citep{2014RAA....14..933Z,2022RAA....22e5017L,2022MNRAS.512.1003Z}.
Thanks to the Fermi Large Area Telescope (Fermi-LAT) for the long-term monitoring of blazars in gamma-ray energy, the first gamma-ray QPO signal of 2.18 years have been found in blazar PG~1553+113 \citep{2015ApJ...813L..41A}. Recently, about 30 blazars showing QPO signals have been discovered by employing multiple techniques. The timescale falls in the range of months-like oscillations of 34.5 days to several years \citep{2015ApJ...813L..41A,2017AA...600A.132S,2017ApJ...835..260Z,2017ApJ...842...10Z,2017ApJ...845...82Z,2018NatCo...9.4599Z,2020MNRAS.499..653K,2020ApJ...896..134P,2022ApJ...929..130W,2023A&A...672A..86R}. The monthly timescale modulation is considered to be the helical structure of the jet, and the periodic signals are generated while the viewing angle of the emission region changes \citep{2018NatCo...9.4599Z}. The presence of a binary black hole systerm is the leading theory for a galaxy with a long-term periodicity with in the order of years QPO behavior \citep{1996ApJ...460..207L,2014ApJ...793L...1S,2021MNRAS.506.3791R}. Thus the Kepler-orbit causes the change in the angle between the jet and observer directions, and the Doppler factor changes cause the periodic phenomenon \citep{2017MNRAS.465..161S,2022ApJ...926L..35O}. Furthermore, the Lense-Thirring precession of the accretion disk can also produce the quasi-periodic phenomena of the annual time scale \citep{2020ApJ...891..120B}. For blazars, the precession of the accretion disk can also affect the jet flow, which leads to the periodic flux variations
\citep{2018MNRAS.474L..81L}. Black hole and accretion disk are the energy sources of jet \citep{1977MNRAS.179..433B,1982MNRAS.199..883B}. Accretion disks may create or destroy QPOs from the jets \citep{2006ApJ...650..749L}. The results of these models are in good agreement with the observations, though further evidence is required in support for these models.

Most of the traditional tools for searching the periodic signals are based on Fourier technique, such as Discrete Fourier Transform \citep{1981AJ.....86..619F,1995AJ....109.1889F}, Lomb–Scargle periodogram \citep{1976Ap&SS..39..447L,1982ApJ...263..835S}, Weighted Wavelet Z-transform \citep{1996AJ....112.1709F}. Fourier techniques are sensitive to sinusoidal variations, but the QPO behaviors of X-ray binaries are considered non-sinusoidal and harmonic \citep{2015MNRAS.446.3516I}. Therefore, the results obtained using these methods may be inaccurate and unreliable. Recently, methods based on Gaussian processes have been proposed \citep{2014ApJ...788...33K,2017AJ....154..220F}. Futhermore, because the telescope observations span a short time, the methods and criteria for confirming a high-significance QPO remain controversial, thus making the QPO analysis in AGNs complicated \citep{2019MNRAS.482.1270C,2021ApJ...907..105Y,2023ApJ...946...52Z}.

S2~0109+22 (GC 0109+224, TXS 0109+224, RXJ0112+2244) is a compact radio-loud AGN at coordinates (J2000) RA=01h12m05.8s and DEC=+22d44m39s. It was first identified as a compact radio object by \citet{1971AJ.....76..980D} and \citet{1972AJ.....77..265P} using National Radio Astronomy Observatory (NRAO) 43-m dish of Green Bank in the 5 GHz. The work \citet{1976ApJS...31..143W} has analyzed the spectra of S2~0109+22 and did not find the absorption or emission lines, and it shows the characteristics of a BL Lac object. \citet{1977AJ.....82..776O} have measured a strong millimeter-wave emission in the 90 GHz and defined it as a BL Lac object. The work \citet{1977AJ.....82..935P} suggest that S2~0109+22 is an intermediate in the type between
the large-amplitude variables (OJ 287) and the smaller-amplitude objects (ON~235, OY~091). The latest redshift estimate is $z\sim0.36$ \citep{2018MNRAS.480..879M}.

The source S2~0109+22 has reportedly shown the possible periodic flux variation in radio band \citet{2004MNRAS.348.1379C}. By employing  University of Michigan Radio Astronomy Observatory (UMRAO) and Metsahovi Radio Observatory data (since 1976 to 2002), they found QPO signals of 3.74 and 4.16 years in 37 and 22 GHz, respectively, 3.43 and 1.8 years in 14.5 GHz, all of results with the significance of \textgreater99\%. Except these, there are no other QPO reported in a different energy band. The search for the QPO behavior, helps us to understand the characteristics and mechanisms of the variability of this blazar. Further, the Fermi-LAT could support the long-term observation in the search for the QPO signals.

In this paper, we searched for the S2~0109+22 period variation by employing the $\gamma$-ray data of Fermi LAT. In Section \ref{sec:obs}, we have introduced the observation and data reduction in the $\gamma$-ray energy bands. In Section \ref{sec:analysis}, we briefly introduce the periodic search methods and give the analysis results. In Section \ref{sec:discuss}, we have discussed the application of the binary black hole model of accretion and lighthouse effect. In Section \ref{sec:conclusion}, we have given the conclusion of our work. 

\section{Observations and Data reduction} \label{sec:obs}

Fermi-LAT can survey the whole sky region with in the energy range of 20 MeV to 300GeV for every three hours \citep{2009ApJ...697.1071A}.
It has made long-term observations of S2~0109+22 (4FGL J0112.1+2245) from 2008-2023 \citep{2020ApJS..247...33A}. We have selected the events file in 2008 August 4 to 2023 January 16 (54682-59974 MJD) from the Fermi Pass 8  database.\footnote{\url{https://fermi.gsfc.nasa.gov/ssc/data/analysis/documentation/Pass8_usage.html}}
The standard criteria for data reduction is the point-source analysis and the analyzing tool is the official software Fermitools 2.0.8.\footnote{\url{https://github.com/fermi-lat/Fermitools-conda/}} Concerning the details of the process, the region of interest (ROI), filter expression and maximum zenith angle value have been used as $15^{\circ}$, (DATA\_QUAL\textgreater0) \& (LAT\_CONFIG=1), $90^{\circ}$, respectively. In the likelihood analysis, we select gll\_iem\_v07.fits and iso\_P8R3\_SOURCE\_V3\_v1.txt as the Galactic diffuse emission model file and the extragalactic isotropic diffuse emission model file, respectively. The instrumental response function is P8R3\_SOURCE\_V3.
Furthermore, the data of S2~0109+22 in 0.1-300 GeV have been selected for the binned maximum-likelihood analysis. We have obtained an average photon flux $(8.37\pm{1.54})\times10^{-8}$ $photons$ $cm^{-2}$ $s^{-1}$ and the test statistic (TS) value is $\sim28872.4$. The corresponding best-fit spectral parameters are $\alpha=2.01\pm{0.01}$, $\beta=0.05\pm0.01$ and $E_{b}=0.769$ GeV with LogParabola model using Fermitools software.

\begin{figure*}[th]
	\figurenum{1}
	\plotone{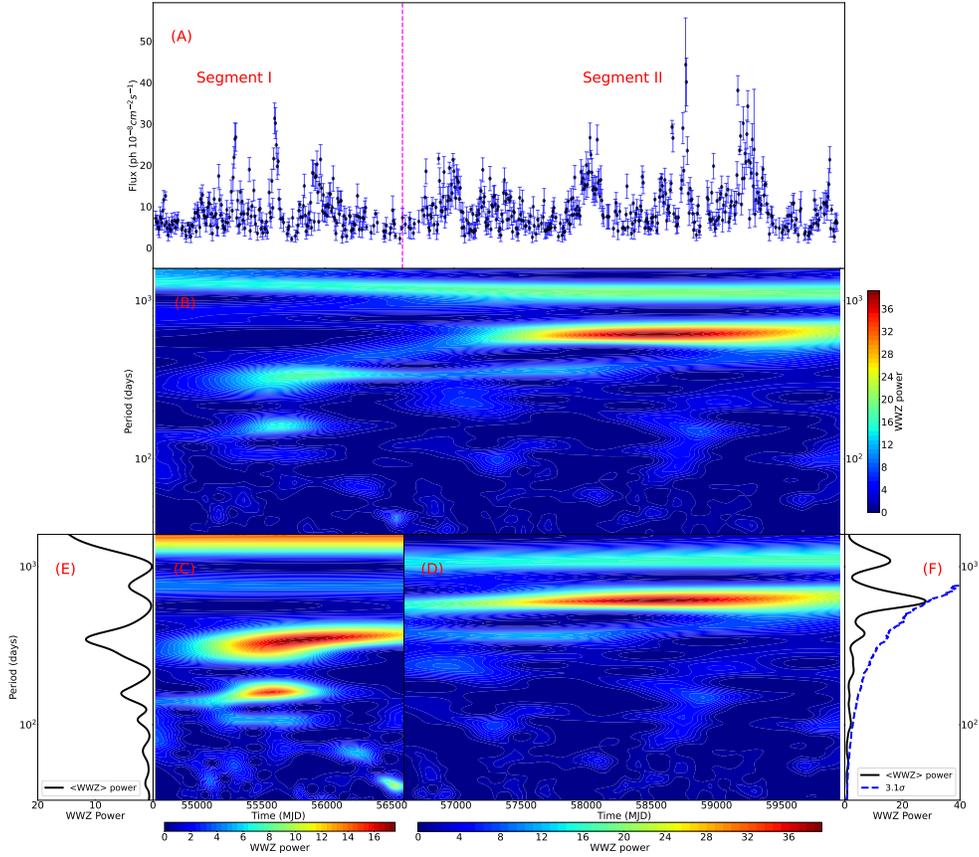}
	\caption{Panel (A) is the Fermi-LAT 0.1-300 GeV $\gamma$-ray light curve with 7-day binned from MJD~54682 to 59974 and the error bar represent $TS>9$ data. The pink dashed line shows the time after which the periodic behavior occurs (Segment II, 56600-59974 MJD). Panel (B) indicates the WWZ 2D analysis results of the complete time span light curve. The results of Segments I and II analysis using WWZ are shown in panels (C) and (D). The corresponding average WWZ power with Segments I and II is expressed by panels (E) and (F), respectively. The black line represents the average WWZ power and the blue dashed line represents the 3.1$\sigma$ significance level.  \label{fig:wwz_result}}
\end{figure*}

Following the determination of the model file with the binned likelihood analysis, we have conducted the unbinned likelihood analysis with 7-day binned. The choice of the 7-day binned has provided the shortest time intervals that can have enough data points to characterize the flux over time, with the maximum likelihood Test Statistic (TS) values exceeding 9 ($\sim$3$\sigma$). The light curve of S2~0109+22 is displayed in top panel of Figure \ref{fig:wwz_result}, where only the flux with the TS\textgreater9 is plotted. To verify the reliability of periodicity, we have adopted the same operation process to extract the 10-day binning and 30-day binning light curves for the analysis.

\section{Data analysis and Results} \label{sec:analysis}
\subsection{Methods}
To study the variability properties of S2~0109+22 in $\gamma$-ray energy, four different methods have been applied to search for the periodic signals in the gamma-ray light curve. Weighted Wavelet Z-transform (WWZ, \citet{1996AJ....112.1709F}) fit the sinusoidal waves that are localized in both the frequency and time domains. It has the advantage of showing the evolution of the peak frequency over time. This method gives the 2D time-frequency power map and time average power spectrum. The Lomb-Scargle periodogram (LSP, \citet{1976Ap&SS..39..447L};\citet{1982ApJ...263..835S}) is a technique similar to the Fourier transform. It can calculate the power spectrum of non-equally spaced sampling time series by assuming a sine function for the periodic component. We employed this method for verifying the periodic behavior found by WWZ. The REDFIT software\footnote{\url{https://www.manfredmudelsee.com/soft/redfit/index.htm}} has been developed by \citet{2002CG.....28..421S}. In this method, the power spectral density (PSD) of time series has been calculated based on the LSP. The red noise level of the PSD has been estimated by fitting a first-order autoregressive process. The advantage of this method is that it can accurately estimate the significance of the broad peak structure in PSD under the red noise background. A light curve folded at a suspicious period should have an approximate periodic shape by using the phase folded light curve. For the Fermi-LAT data, we have used the phase-resolved binned likelihood analysis with the obtained period.

We have used the $baluev$ method given by \citet{2008MNRAS.385.1279B} to describe the False Alarm Level (FAL) of the peak given by LSP, and to estimate the FAL of the peak in PSD (see the red dashed line in Figure \ref{fig:lsp}). To eliminate the noise from the false periodic signal, we have employed the light curve simulation software\footnote{\url{https://github.com/samconnolly/DELightcurveSimulation}} developed by \citet{2013MNRAS.433..907E} to estimate the significance of the result. This technique is based on the assumption that the PSD of AGN is a red noise spectrum \citep{1995A&A...300..707T}, henceforth giving the simulated light curve by fitting the PSD and probability density function (PDF) of the light curves. We have simulated $10^{5}$ light curves for the LSP method and WWZ method to calculate the significance level of these results.
\subsection{Results}
WWZ can reveal the dynamic fluctuation of the power spectrum over time. The results of WWZ analysis on the data are shown in Figure \ref{fig:wwz_result}. We have shown the 7~days~bin$^{-1}$ complete time span light curve (54682-59974 MJD) and the corresponding WWZ power diagram, in the panels (A) and (B) of Figure \ref{fig:wwz_result}. The power increase can be seen in the Segment II (56600-59974 MJD), for a time scale of $\sim600$ days. To analyse the periodic behavior accurately, we have analysed Segment I (54682-56600 MJD) and Segment II with WWZ. Panels (D) and (F) show the WWZ 2D map and the average power of Segment II, respectively. We have obtained $604\pm{78}$ days of periodic signals with $3.1\sigma$ significance level about average WWZ power. The uncertainty range of the period was the full width at half maxima (FWHM) of the peak fitted by a Gaussian function. We also used the same operations for Segment I (see in panel (C) and panel (E)). Although it shows a short periodic behavior, the significance was less than $2\sigma$. We did not consider this transient periodic behavior in the subsequent discussions.

\begin{figure}[t]
	\figurenum{2}
	\plotone{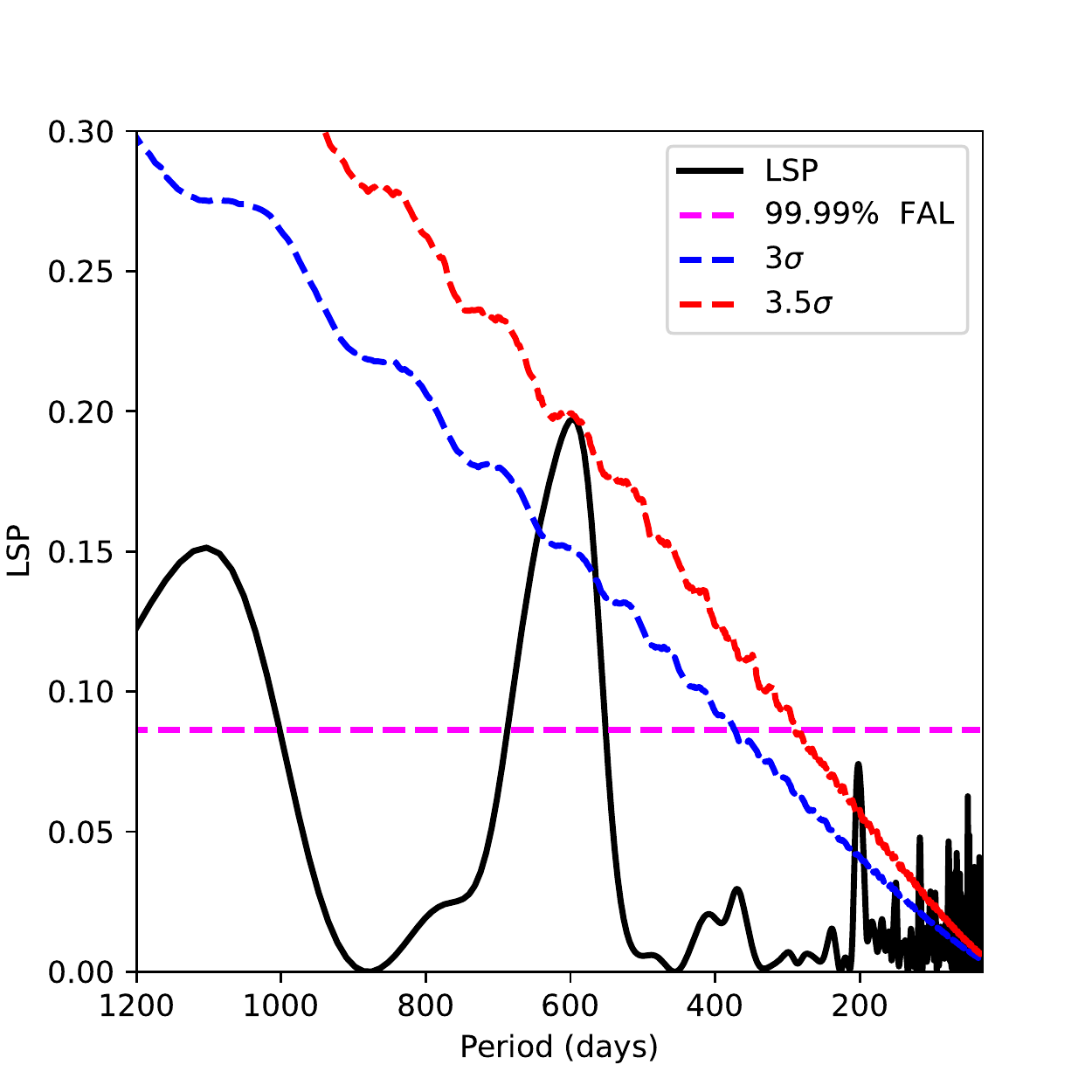}
	\caption{Possible period obtained from LSP in Segment II. The black line is the LSP power, in which a broad peak with peak value of $596\pm{62}$ days. The dashed red, blue, and pink lines correspond to the 99.99\% FAL, 3 $\sigma$ and 3.5 $\sigma$ significance levels, respectively. \label{fig:lsp}}
\end{figure}

We have found a period of $596\pm{62}$ days (see in Figure \ref{fig:lsp}) from the LSP analysis results of MJD 56600 to 59974 (Segment II). The peak exceeds 99.99\% FAL and achieve 3.5 $\sigma$ significance.

\begin{figure}[t]
	\figurenum{3}
	\plotone{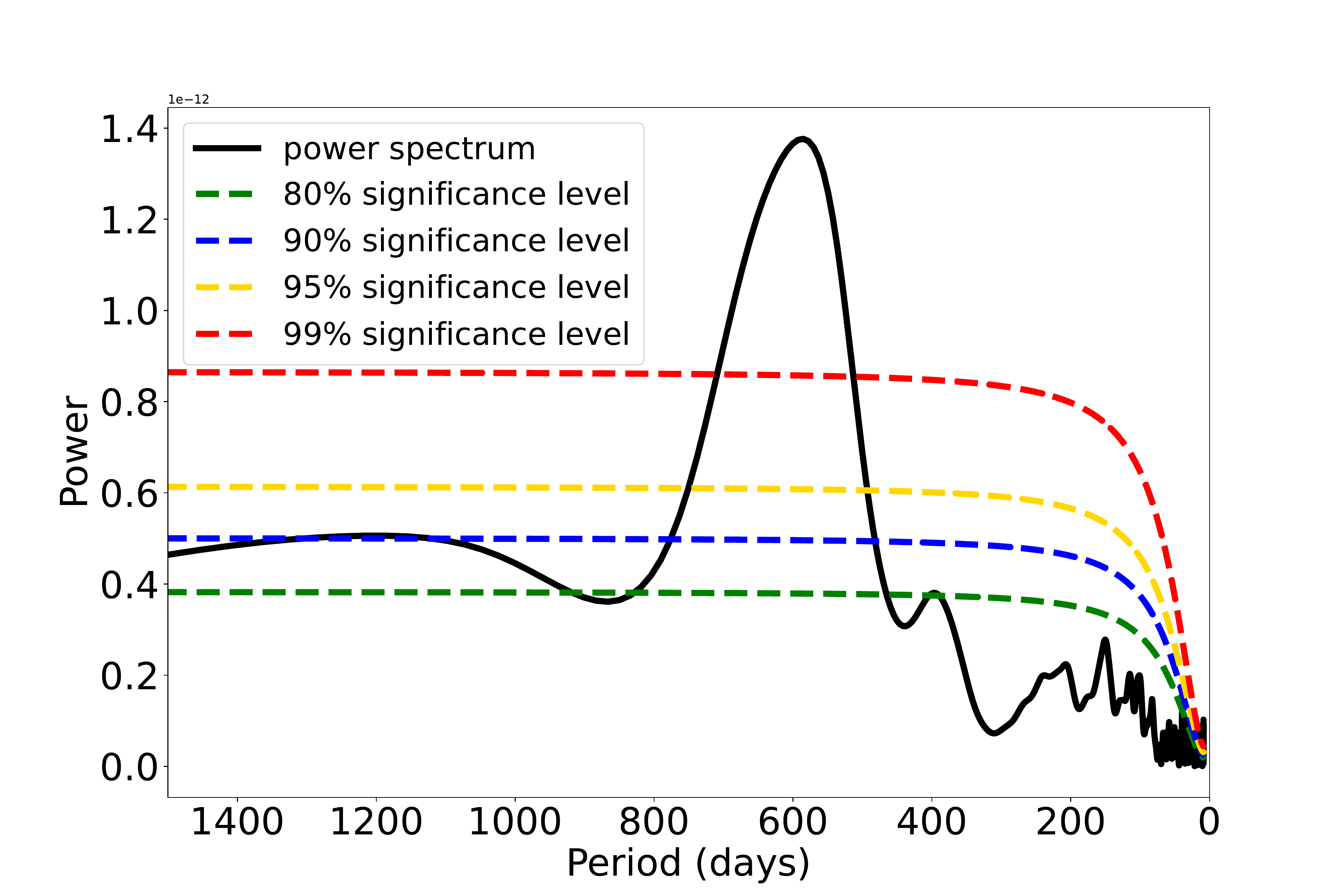}
	\caption{Results from analysis of the REDFIT program. The black line is the PSD calculated by REDFIT, the peak value is $588\pm{80}$ days. The dashed lines with green, blue, yellow and red colors indicate the significance level by estimating the red noise background. \label{fig:redfit}}
\end{figure}

The result of analysis from the Segment II using REDFIT program is shown in Figure \ref{fig:redfit}. The peak was well above 99\% significance level at $588\pm{80}$ days. Although the period was slightly short than those of the previous two methods, they were within the allowable error range of the value.

\begin{figure}[t]
	\figurenum{4}
	\plotone{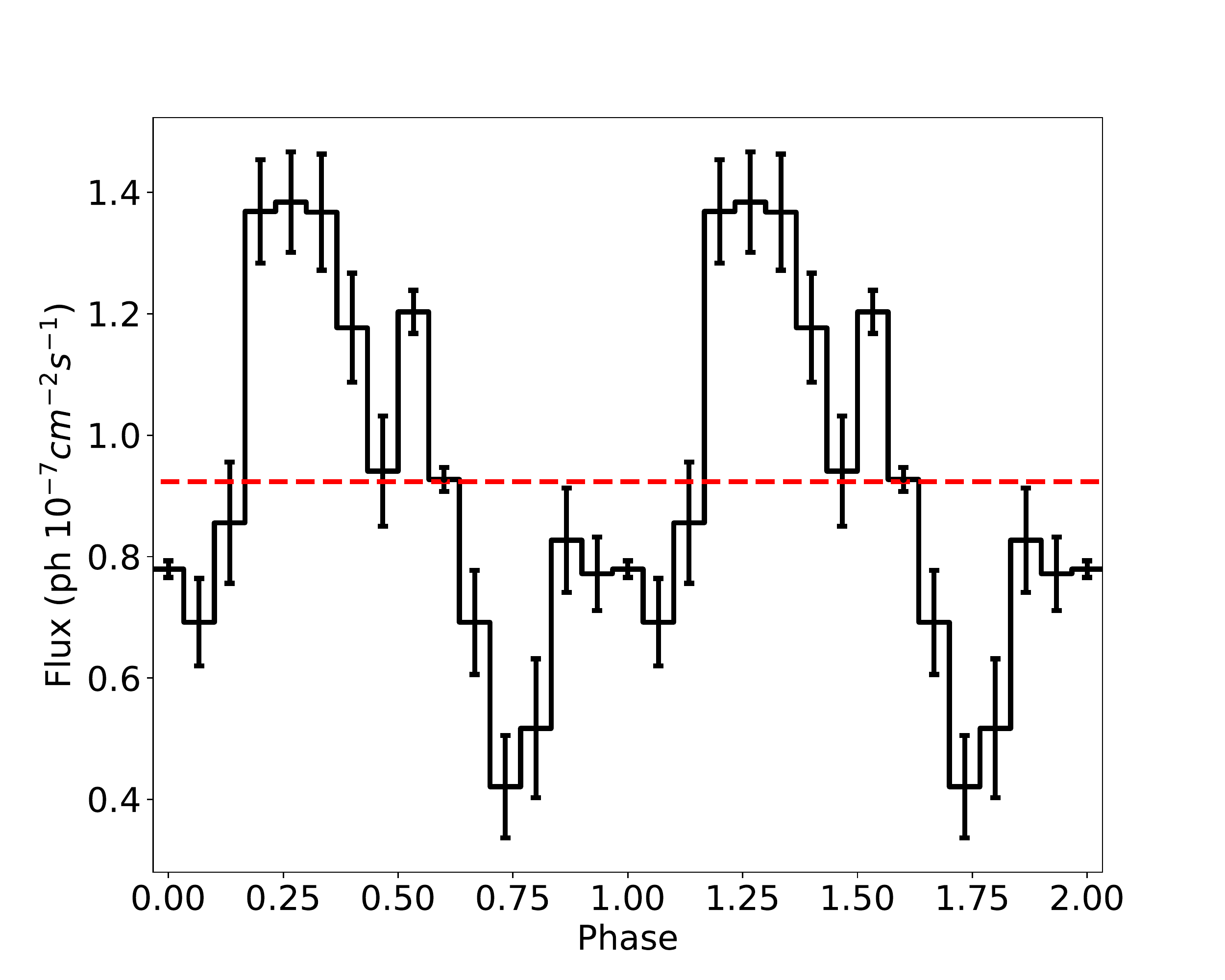}
	\caption{Phase folded light curve with period of 600 days existed in Segment II, from 56685 to 59685 MJD. Two clarified cycles are shown. The dashed red line is the mean flux.\label{fig:fold}}
\end{figure}

To confirm the reliability of our results, a phase folded technique has been employed for validation. Figure \ref{fig:fold} shows, the significant periodic shape when we fold it over a period of 600 days from 56685 to 59685 MJD.

Four methods have been found to have a significant quasi-periodic behavior with $\sim600$ days at 56600-59974 MJD, with the LSP giving a significance level of about $3.5\sigma$ and well above $99.99\%$ FAL. We list the results of search for the periodic signals using four methods, in Table \ref{tab:result}. We have also analyzed the data of 10 days bin$^{-1}$ and 30 days bin$^{-1}$ of S2 0109+22. The quasi-periodic behavior of $\sim600$ days was found to be evident after different time cadence.

\begin{deluxetable*}{ccccc}[t]
	\tablenum{1}
	\tablecaption{Results of the QPO analysis with 4 methods \label{tab:result}}
	\tablewidth{0pt}
	\tablehead{
		Method & UMRAO 14.5 GHz (1980-2002) & \colhead{Fermi-LAT $\gamma$-ray (2014-2022)} & \colhead{Significance level}
	}
	\startdata
	WWZ & & $604\pm{78}$ days & $3.1\sigma$\\
	LSP	& $\sim657$ days & $596\pm{62}$ days & $3.5\sigma$\\
	REDFIT & & $588\pm{80}$ days & $\textgreater 99\%$\\
	phase folded light curve & & $\sim600$ days & \\
	\enddata
	\tablecomments{The result of UMRAO 14.5 GHz is taken from \citet{2004MNRAS.348.1379C} and the significance value is \textgreater99\%.}
\end{deluxetable*}

\section{Discussion} \label{sec:discuss}
The long-term $\gamma$-ray data have been analyzed using four different techniques to look for periodicity. These methods delivered consistent results. Previously, \citet{2004MNRAS.348.1379C} reported that this source had a radio quasi-periodic modulation of $\sim657$ days with a \textgreater99\% confidence level. In our work, the periodicity of the two energy bands is consistent while considering the error range, which strongly supports our finding. Figures \ref{fig:wwz_result} (F) and \ref{fig:lsp} show that an  another peak has been detected at a location of approximately 1100 days with a FAL above 99.99\%, though the simulated light curves clearly gave a result that was below the 3$\sigma$ significance level. We generally considered signals that exceeded both of the significance estimation methods are reliable. This may be because the analysis methods were based on the Fourier transform and decomposed the original signal into a sum of frequency components, including harmonic frequencies \citep{2022ApJ...929..130W}. Therefore, in the following discussion, we did not consider it to be a real signal, but rather a frequency doubling of the detected QPO. Consequently, we confirmed that there was an obvious periodic behavior for BL Lac S2~0109+22 in $\gamma$-ray with $3.5\sigma$ significance level of $\sim600$ days (1.64 years). 

\begin{figure}[h]
	\figurenum{5}
	\plotone{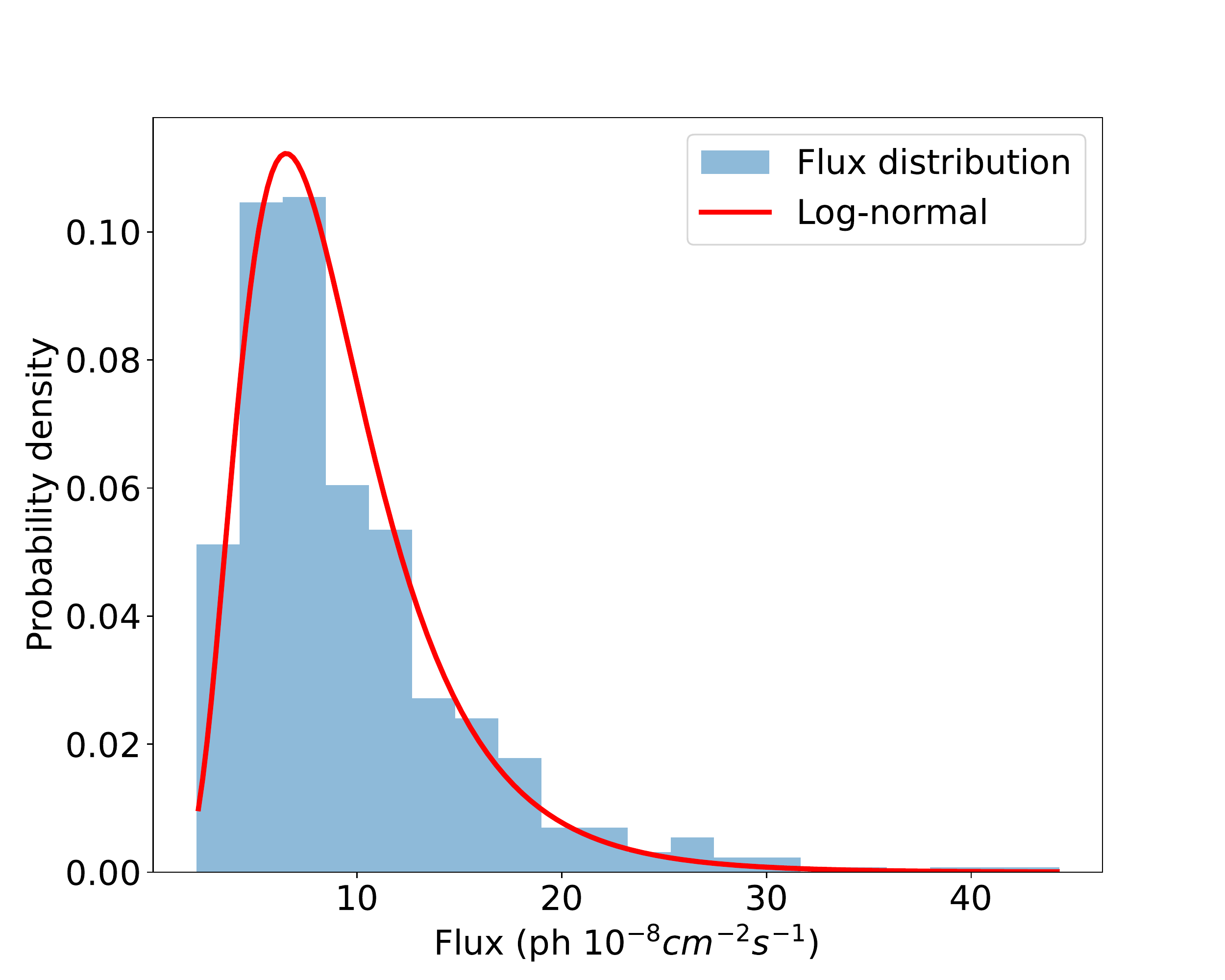}
	\caption{Histogram shown as the flux PDF of $\gamma$-ray light curve of S2 0109+22. The red line is the lognormal distribution fit with the parameters $s=0.48$ and $scale=8.26$ by the Python package \textbf{scipy.stats.lognorm}. \label{fig:pdf}}
\end{figure}

In Segment I, we found no evidence of a periodic activity. The $\gamma$-ray light curve PDF of S2 0109+22 exhibitsed a significant lognormal distribution shape (see in Figure \ref{fig:pdf}). Lognormal distribution will typically be visible in the PDF of X-ray binaries with the accretion disk emission dominating them \citep{2009ApJ...697L.167G}.  The accretion disk and the jet radiation may have significant connections \citep{2022A&A...664A.166R,2022ApJ...936..146X,2022ApJ...930..157Z}.
The accretion disk's instability caused turbulence disturbance to emerge at random in various radii. Such an effect will spread from the outside to the inside of the disk. Once it hits the innermost region, it may impact the soft photon field and electron injection into the jet, which in turn effect the emission of the jet \citep{1997MNRAS.292..679L,2006MNRAS.367..801A}. The extremely high bulk Lorentz factor of the jet had the potential to enhance the effect of the accretion disk and subsequently may cause significant variations in $\gamma$-ray radiation. This scenario could be a factor in the periodic signals disappearing from Segment I. 

\begin{figure*}[t]
	\figurenum{6}
	\plotone{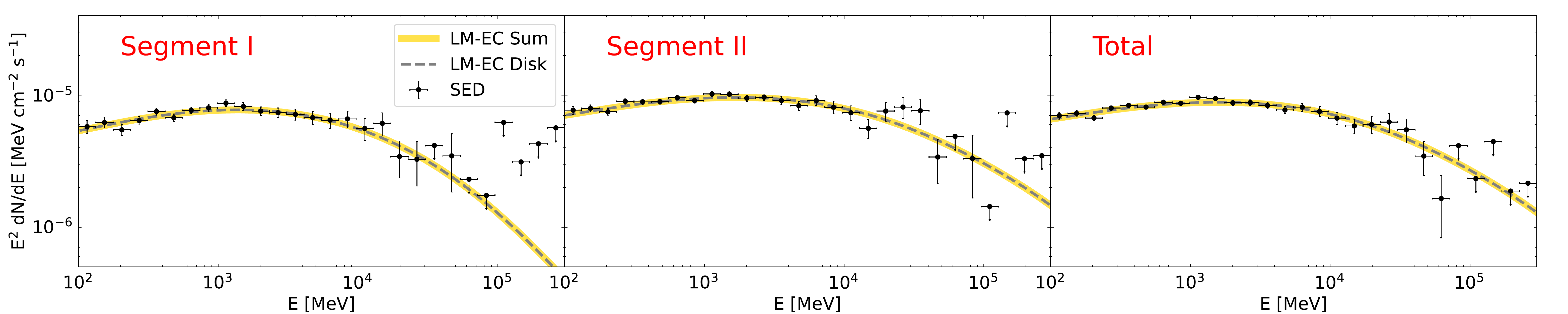}
	\caption{Segments I and II, and Total SED are represented by balck error bars for the left, middle, and right panels, respectively. The gray dashed and yellow lines shown the SED fitting results of the accretion disk contribution and the sum of each component contributions in LM, respectively. \label{fig:sed_fit}}
\end{figure*}

To investigate scenario, we have applied the Leptonic Model (LM) to explore possible evidence for this \citep{1992ApJ...397L...5M,1994ApJ...421..153S,2006A&A...448..861M,2011ApJ...739...66T,2013ApJ...768...54B}. Considering that the accretion disk provides the external seed photons, we use the External Compton (EC) model to fit the $\gamma$-ray Spectral Energy Distribution (SED) of this source \citep{1995ApJ...441...79B,1998A&A...336..123B}. A powerful SED fitting software--JetSet\footnote{\url{https://github.com/andreatramacere/jetset}} \citep{2020ascl.soft09001T}, has been employed to fit the SED of Segment I, Segment II and Total, respectively. During the fitting process, we had to freeze certain parameters to the typical values found in blazars owing to the lack of multiwavelength data. We found that the LogParabola electron spectral was the most suitable one. In Figure \ref{fig:sed_fit}, there was an obvious spectral hardening in Segment II compared to Segment I. Under the broadly similar radiation region size ($\sim1.4\times10^{15}$ cm), the electron density increases from Segment I to Segment II (from $1.99\pm0.46$ to $5.64\pm{0.14}$, with units of $10^4$ cm$^{-3}$ as shown in Table \ref{tab:sed_params}). These parameters exhibit a similar order of magnitude as the constraints on physical parameters of high-energy radiation in blazars by \citet{1998ApJ...509..608T}. The accretion disk could inject electrons into the bottom of the jet \citep{2020Univ....6...99R}. Consequently, the electrons in the radiant region that were close to the black hole may be affected by the accretion disk. If the changes in the electrons proceeded from perturbations in the accretion disk, then the seed photons were affected. The periodic phenomenon in Segment II gradually gained strength over time, which may ensue from the dissipating perturbations (see in Figure \ref{fig:wwz_result} (A)). We performed a simple quantitative analysis of this scenario, because further study was beyond the scope of this work and requires multiwavelength SED.

The long-scale quasi-periodic behavior of blazars is often interpreted as the supermassive binary black hole (BBH) systems \citep{1988ApJ...325..628S,1996ApJ...460..207L,2007A&A...462..547F,2014ApJ...793L...1S}. This scenario consists of two parts, viz., the accretion and jet lighthouse models. In the accretion model, when the secondary black hole is close to the primary black hole owing to orbital motion, the accretion rate increases significantly, hence the flux will rise periodically \citep{2022ApJ...929..130W}. The jet lighthouse model considers that the periodic change of the observation angle brings about the change of the Doppler factor, thus generating the periodicity of the flux \citep{1998MNRAS.293L..13V,2007ChJAA...7..364Q}. These models provide a geometric explanation for periodic behavior without relying on specific physical parameters.

First, we needed to calculate the periodic expansion due to the redshift effect which was the formula $T_{sou}=T_{obs}/(1+z)$. Then we applied the first model described above to estimate the distance between the two black holes, and considered the motion of BBH in the case of Kepler orbits, given as:
\begin{equation}
	P_{orbital}^{2}=\frac{4\pi^{2}d^{3}}{G(M+m)}
\end{equation}
where $d$ represents the distance between two black holes, and $M$ and $m$ is the primary black hole mass and secondary black hole mass, respectively. $G$ represents the universal gravitational constant. By substituting the redshift $z=0.36$ of this source \citep{2018MNRAS.480..879M}. Furthermore, owing to the periodic flux from the orbital effects, we set $P_{orbital}=T_{sou}=1.21$ years. The average black hole mass of blazar is $\bar{M}=10^{8.6}M_{\sun}$ given by \citet{2021MNRAS.506.3791R}. Furthermore, by assuming $m/M\sim0.001$ by \citet{2022ApJ...929..130W}, we calculated the distance between two black holes as $d=1.25\times10^{16}~cm$.

In the lighthouse model, the periodicity originated from the orbital effect of the BBH, which affected the emission toward to the observer. Thus, the orbital period is equal to the QPO existing in the light curve. Generally, the light curve showed a double flare peak structure \citep{2007ChJAA...7..364Q,2021ApJS..253...10F,2022ApJ...929..130W}. This suggested a possible binary black hole jet structure. Hence, we assume that this is a binary black hole system with a dual jet. Furthermore, we set component-1 for the primary black hole and component-2 for the secondary black hole. The observation angle of two jets changing with time can be \citet{2007ChJAA...7..364Q} expressed as:
\begin{equation}
	\cos\theta_{1}(t)=\sin\psi_{1}\cos(\omega t+\phi_{1})\sin i+\cos\psi_{1}\cos i 	
\end{equation}
\begin{equation}
	\cos\theta_{2}(t)=\sin\psi_{2}\cos(\omega t+\phi_{2})\sin i+\cos\psi_{2}\cos i	
\end{equation}
where $\phi_{1}$ and $\phi_{2}$ are the azimuths angles between the observer and jet axis of two components on the orbital plane. The two angles are expressed as the initial phase of the two periodic jet radiation. The difference of those is the interval between the two flares. According to the works \citep{2007ChJAA...7..364Q,2021ApJS..253...10F,2022ApJ...929..130W}, we set $t=0$, the azimuths angle between the observer and component-1 is $\phi_{1}=80^{\circ}$ and $\phi_{2}=240^{\circ}$ for component-2, according to the shape of the flux variability. The angles between jet axis and orbital axis for component-1 and component-2 were $\psi_{1}$ and $\psi_{2}$, respectively. Furthermore, $i$ was set as the sight direction with the orbital axis and the parameter $\omega$ was calculated based on ($2\pi/T_{obs}$). Therefore, we obtained the representation of the variable Doppler factors ($\delta_{1}$ and $\delta_{2}$):
\begin{equation}
\begin{split}
	\delta_{1}(t)=[\Gamma_{1}(1-\beta_{1}\cos\theta_{1}(t))]^{-1} \\
	\delta_{2}(t)=[\Gamma_{2}(1-\beta_{2}\cos\theta_{2}(t))]^{-1}
\end{split}
\end{equation}
where $\Gamma=(1-\beta^{2})^{-1/2}$ is the Lorentz factor and, $\beta$ is the bulk velocity of the jet in the unit of light speed. The apparent velocity ($\beta_{app1}$ and $\beta_{app2}$) are \citep{2019SCPMA..6229811X,2022ApJ...929..130W}:
\begin{equation}
\begin{split}
	\beta_{app1}^{2}(t)=-\delta_{1}(t)^{2}-1+2\delta_{1}(t)\Gamma_{1} \\
	\beta_{app2}^{2}(t)=-\delta_{2}(t)^{2}-1+2\delta_{2}(t)\Gamma_{2} 
\end{split}
\end{equation}\\
The changed flux caused by the change of Doppler factors are given as:
\begin{equation}
	f_{1}(t)=f_{b1}\delta_{1}(t)^{4},~f_{2}(t)=f_{b2}\delta_{2}(t)^{4}
\end{equation}
where $f_{b1}$ and $f_{b2}$ are the normalization constant of the light curve fitting.

\begin{deluxetable*}{cccccc}[t]
	\tablenum{2}
	\tablecaption{Parameters for the LM Model. \label{tab:sed_params}}
	\tablewidth{0pt}
	\tablehead{
	\colhead{Parameter} & \colhead{Segment I} & \colhead{Segment II} & \colhead{Total} & \colhead{Reference} & \\}
	\startdata
	Electron Density ($1/cm^{3}$) & $(1.99\pm{0.46})\times10^{4} $ & $(5.64\pm{0.14})\times10^{4}$ & $(5.27\pm{0.74})\times10^{4}$ & ... & \\ 
	Radiation Region Size ($cm$) & $(1.51\pm{0.20})\times10^{15} $ & $(1.30\pm{0.21})\times10^{15}$ & $(1.52\pm{0.13})\times10^{15}$ & ... & \\ 
	Electron Spectral Slope & $2.34\pm{0.07}$ & $2.45\pm{0.04}$& $2.46\pm{0.02}$& ... & \\
	Electron Spectral Curvature & $(1.60\pm{0.53})\times10^{-1} $ & $(-2.17\pm{0.07})\times10^{-1}$ & $(-2.09\pm{0.03})\times10^{-1}$& ... & \\
	Electron Spectral cut-off $\Gamma$ & $(1.90\pm{0.56})\times10^{2} $ & $(8.27\pm{1.33})\times10^{4}$ & $(4.51\pm{0.40})\times10^{4}$& ... & \\
	Radiation position$^{\textbf{*}}$ ($cm$) & & $5\times10^{14}$ & & 1 & \\
	$\Gamma^{\textbf{*}}$ & & 35 & & 2 & \\
	$M_{BH}^{\textbf{*}}$ & & $10^{8.6}M_{\sun}$ & & 3 & \\
	$z^{\textbf{*}}$ & & 0.36 & & 4 & \\
	$L_{disk}^{\textbf{*}}$ ($erg/s$) & & $1\times10^{45}$ & & 5 & \\
	$\dot{M}/\dot{M}_{edd}^{\textbf{*}}$ & & $0.12$ & & 5 & \\
    Reduced $\chi^{2}$ & 0.88 & 1.25 & 1.53 & ... & \\	
	\enddata
	\tablecomments{References: (1) \citet{2009AA...501..879T}, (2) \citet{2022ApJ...929..130W}, (3) \citet{2021MNRAS.506.3791R}, (4) \citet{2018MNRAS.480..879M}, (5) \citet{2003APh....18..377D}. The superscript " \textbf{*} " indicates the frozen parameter in fit.}
\end{deluxetable*}

\begin{figure}[t]
	\figurenum{7}
	\plotone{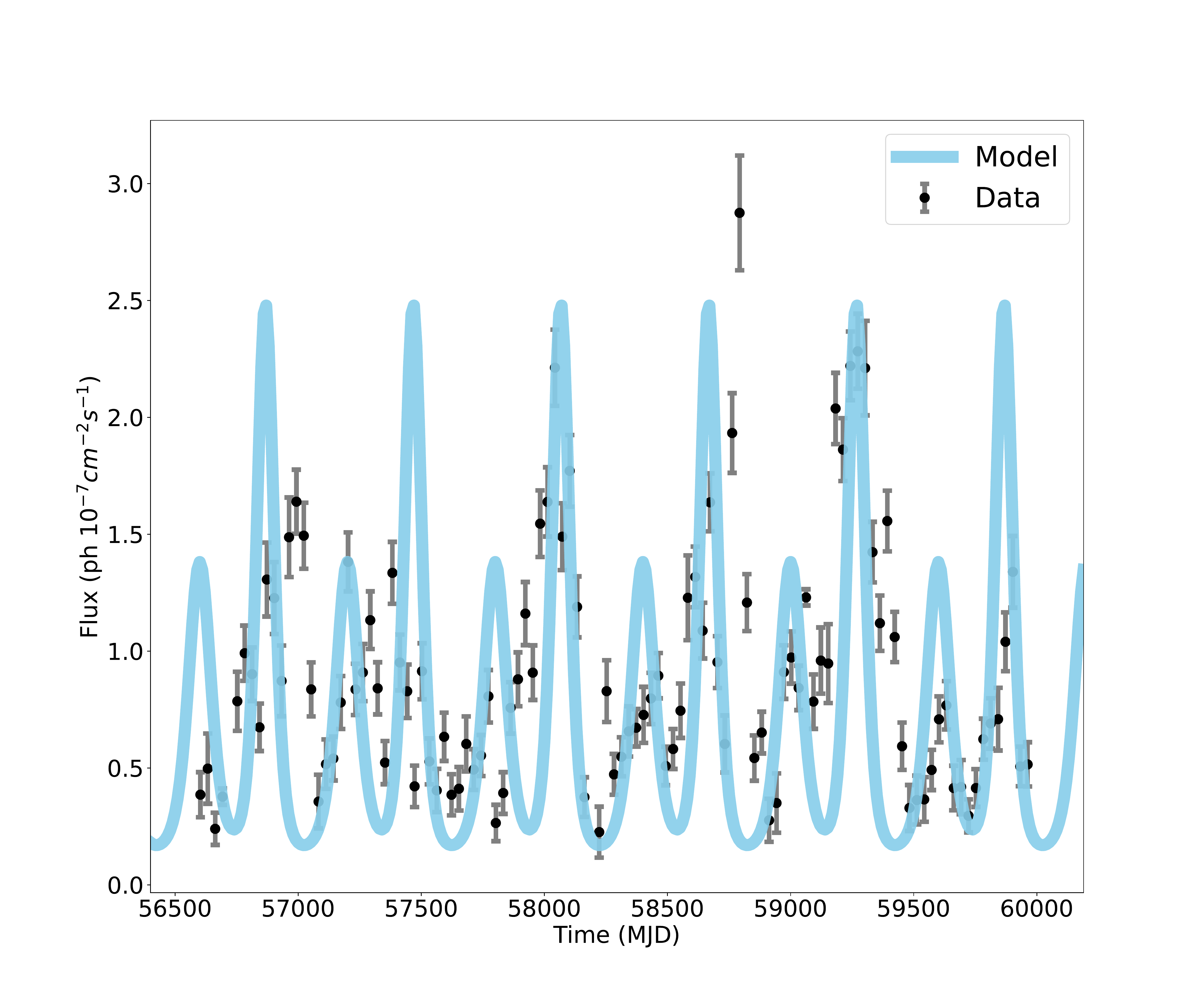}
	\caption{Fitting the lighthouse model. The light blue line indicates the results of lighthouse model fitting, and the gray error bar shows the 30 days bin$^{-1}$ data from Fermi-LAT.\label{fig:fit_result}}
\end{figure}

\begin{figure*}[t]
	\figurenum{8}
	\gridline{\fig{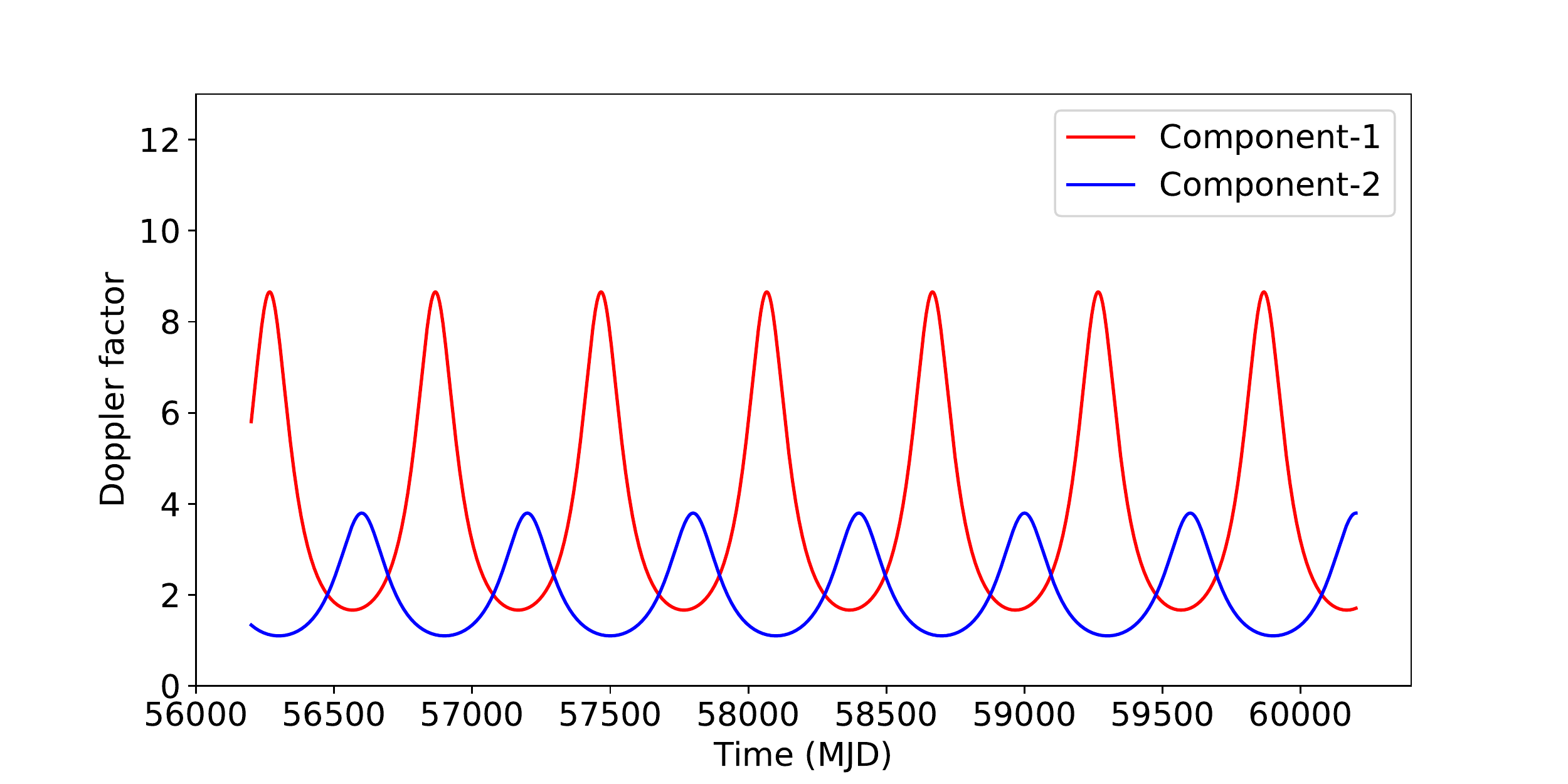}{0.5\textwidth}{}
		\fig{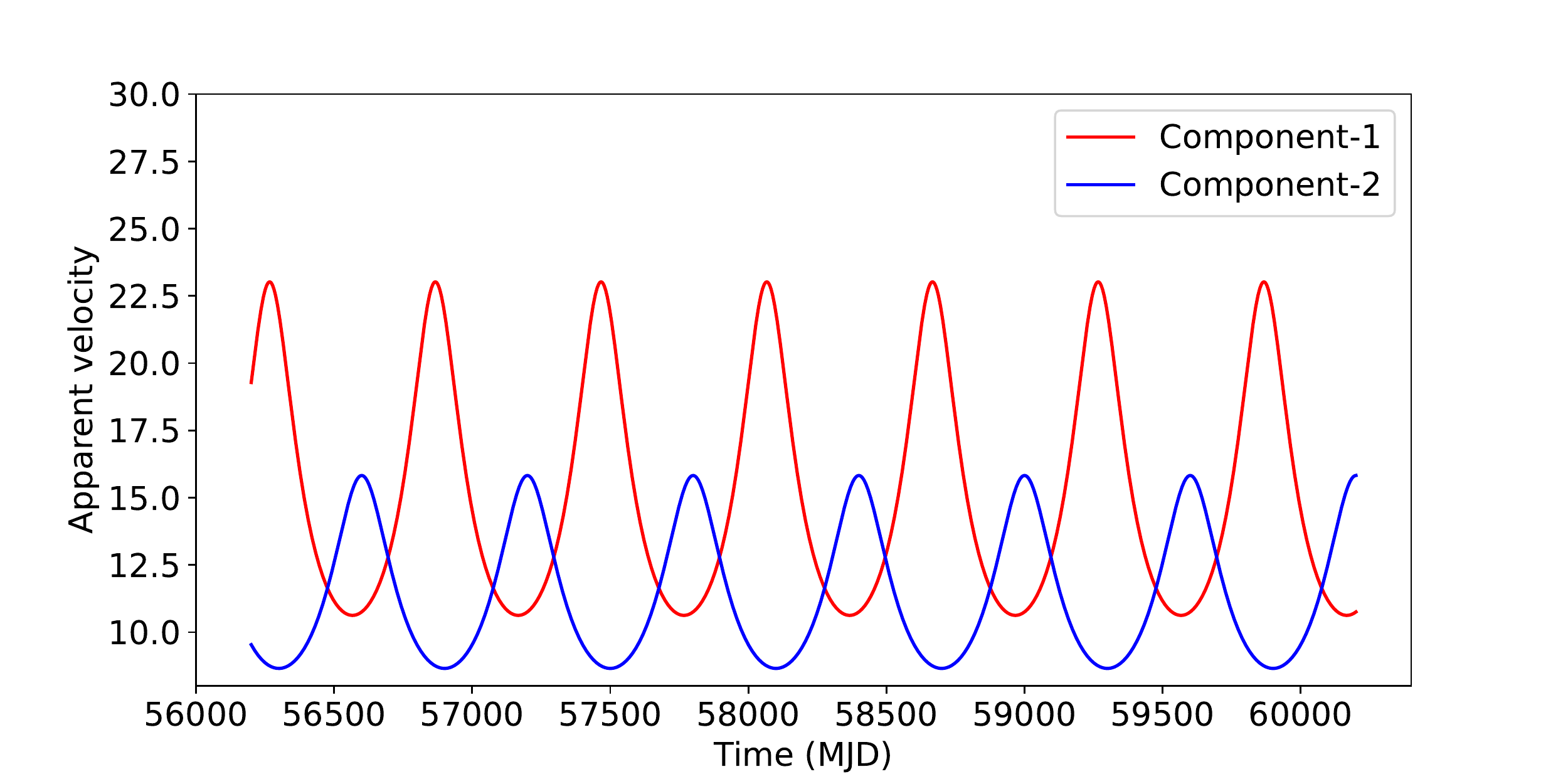}{0.5\textwidth}{}
	}
	\caption{The top and bottom panel indicate the changes of Doppler factor and apparent velocity, respectively. The red and blue lines show the primary black hole (component-1) and the secondary black hole (component-2), respectively. \label{fig:par_result}}
\end{figure*}

\begin{deluxetable}{ccc}[t]
	\tablenum{3}
	\tablecaption{Parameters for the Lighthouse Model. \label{tab:fit}}
	\tablewidth{0pt}
	\tablehead{
		\colhead{Parameter} & \colhead{Value} &\\}
	\startdata
	$\Gamma^{\textbf{*}}$  & 35   &  \\
	$T_{obs}$ & 600 days & \\
	$\psi_{1}$ & $7.43\pm{0.08}^{\circ}$ & \\
	$\phi_{1}^{\textbf{*}}$ & $80^{\circ}$ & \\
	$\psi_{2}$ & $9.91\pm{0.03}^{\circ}$ & \\
	$\phi_{2}^{\textbf{*}}$ & $240^{\circ}$ & \\
	$i$ & $3.07\pm{0.08}^{\circ}$ & \\
	$f_{b1}$ & $(0.42\pm{0.08})\times10^{-3}$ & \\
	$f_{b2}$ & $(0.49\pm{0.02})\times10^{-2}$ & \\
	quiescent level & $(0.12\pm{0.002})\times10^{-7}~ph~cm^{-2}~s^{-1}$ & \\
	\enddata
	\tablecomments{$\Gamma$ is refer to \citet{2022ApJ...929..130W}. The superscript " \textbf{*} " indicates the frozen parameter in fit.}
\end{deluxetable}

Following the determination the QPO timescale, the lighthouse model has been used to fit the light curve. We used the \textbf{iminuit} program \footnote{\url{https://iminuit.readthedocs.io/en/stable/about.html}} \citep{iminuit} for maximum likelihood fitting of the data. The fitting parameters are shown in Table \ref{tab:fit}. The reduced $\chi^{2}$ of the model is 1.67 obtained by \textbf{iminuit}. The model agreed the overall trend of the curve, and the points outside the model were proximate to the model curve without considerable deviation (see in Figure \ref{fig:fit_result}). As mentioned previously, the jet radiation was affected by various factors such as internal instability of accretion disk, hence it was difficult for a simple BBH jet model to fit perfectly. We also gave the law of Doppler factors ($\delta_{1}(t)$,~$\delta_{2}(t)$) and apparent velocities ($\beta_{app1}$,~$\beta_{app2}$) changing with time in the Figure \ref{fig:par_result}. The periodic change of Doppler factor caused the formation of quasi-periodic signal and, in turn, the periodic change of apparent velocitiy.

The best fitting of the Doppler factor range of component-1 and component-2 were $1.67~\textless~\delta_{1}~\textless~8.66$ and $1.10~\textless~\delta_{2}~\textless~3.80$, respectively. \citet{2013RAA....13..259F} have estimated the Doppler factors of 138 Fermi blazars, where the result for S2~0109+22 is $\delta_{\gamma}=2.59$. Our model fitting parameter of Doppler factor was compatible with this result. We also obtained the apparent velocity range with $10.62~\textless~\beta_{app1}~\textless~23.02$ for component-1 and $8.65~\textless~\beta_{app2}~\textless~15.83$ for component-2. \citet{2019SCPMA..6229811X} have suggested that the latest range of superluminal velocity was $0.53~\textless~\beta_{app}~\textless~34.80$, and our result fell within this range.

Because the $\gamma$-ray flux distribution was lognormal, we speculated that the disturbance of the accretion disk to the jet may caused the periodic disappearance in the Segment I. Periodic accretion model and binary black hole lighthouse model have been employed to explain this quasi-periodic phenomenon. The lighthouse model could not fit all the data points perfectly. Therefore, we believed that this quasi-periodic phenomenon was dominated by the variable Doppler factor caused by dual jets, whereas the periodic accretion model was secondary.

\section{Conclusion} \label{sec:conclusion}
We have analyzed, the Fermi-LAT data of Blazar S2~0109+22 from 2008 to 2022. Four different methods have been employed to search for the periodicity. We have found a $\gamma$-ray QPO behavior of $\sim600$ days with a significance $\sim3.5$$\sigma$ from MJD 56600 to 59974. Within 54682-56600 MJD, we suggested that turbulence from the accretion disk affected the jet radiation and disrupts the periodic variability. The distance between two black holes, according to our calculations, was $\sim1.25\times10^{16}~cm$. The total range of the two components of Doppler factor in the lighthouse model was $1.10~\textless~\delta~\textless~8.66$. We have also obtained the total apparent velocity range with $8.65~\textless~\beta_{app}~\textless~23.02$, which indicated that this was a superluminal source. Therefore, it was possible that Blazar S2 0109+22 was a binary black hole system with a dual jet structure and a periodic accretion rate. Further credible data will be needed to support this quasi-periodic behavior, and we will continue tracking it in a future work.

\begin{acknowledgments}
\centerline{Acknowledgments}
We thank the anonymous referee for providing constructive
comments and suggestions. This work was partly supported by the National Science Foundation of China (12263007 and 12233006), the High-level talent support program of Yunnan Province. 
\end{acknowledgments}

\facilities{\textit{Fermi} (LAT)}
\software{Astropy \citep{2022ApJ...935..167A},  Fermitools, REDFIT \citep{2002CG.....28..421S}, iminuit \citep{iminuit}, JetSet \citep{2020ascl.soft09001T}
}
\newpage
\bibliography{S20109}{}
\bibliographystyle{aasjournal}

\end{document}